\documentclass[preprint,5p]{elsarticle}

\DeclareUnicodeCharacter{0301}{\'{e}}

\usepackage{amssymb}
\usepackage{amsmath}
\usepackage{hyperref}
\usepackage{float}
\hypersetup{colorlinks=true,
            linkcolor=blue,
            filecolor=magenta,      
            urlcolor=cyan,
            pdftitle={An Example},
            pdfpagemode=FullScreen,}
\usepackage{tabularx}
\usepackage[section]{placeins}

\usepackage{lineno} 
\usepackage{float}
\usepackage{comment}

\journal{Powder Technology}

\begin{document}
\begin{frontmatter}



\title{The effect of baffles on the hydrodynamics of a gas-solid fluidized bed studied using real-time magnetic resonance imaging}



\author{Hannah S. Buchholz}
\author{Daniel L. Brummerloh} 
\author{Stefan Benders}
\author{Alexander Penn}

\address{Institute of Process Imaging, Hamburg University of Technology, Hamburg, Germany}


\begin{abstract}
Understanding and predicting the hydrodynamics of gas bubbles and particle-laden phase in fluidized beds is essential for the successful design and efficient operation of this type of reactor. In this work, we used real-time magnetic resonance imaging (MRI) to investigate the effect of baffles on gas bubble behavior and particle motion in a fluidized bed model with an inner diameter of 190\,mm. MRI time series of the local particle density and velocity were acquired and used to study the size, number, and shape of gas bubbles as well as the motion of the particle phase. The superficial gas velocity was varied between 1 and 2\,$U_\mathrm{mf}$. We found that baffles decreased the average equivalent bubble diameter with a simultaneous increase in the total number of bubbles. Moreover, baffles promoted bubble splitting and the formation of air cushions below the baffles and decreased the average particle velocity and acceleration in the bed. For two of the three investigated baffle types, the particle velocity distribution became wider compared to the bed without internal.
\end{abstract}


\end{frontmatter}


\section{Introduction}\label{Introduction}
Fluidized beds are used in various industrial applications, including the drying, combustion, and coating of granular media, or in highly endo- or exothermic reactions. Compared to most other reactor types, fluidized bed reactors exhibit a higher heat and mass transfer between the involved phases. Fluidized beds can operate in different flow regimes depending on the Geldart classification \cite{Geldart.1973} of the used particles and the superficial gas velocity. For superficial gas velocities below the minimum fluidization velocity, gas flows through the porous structure of the fixed bed. Above the minimum fluidization velocity, large instabilities cause up-streaming gas bubble formation in the fluidized bed \cite{Cuenca.1995}. The occurring gas bubbles with small diameters in fluidized beds enhance the particle mixing \cite{Amiri.2016}. Bubbles with large diameters up to the bed diameter on the other hand decrease the mass and heat transfer in the system and can reduce the contact between the gas and solid phase. Hence, it is important to control the formation and size of gas bubbles within fluidized beds in order to maximize process efficiency. To influence the bubble size and formation by altering the bed geometry, internals, such as baffles, tubes, or packings, can be added to the fluidized bed. From previous investigation it is known that baffles in fluidized beds decrease the average bubble diameter \cite{Dutta.1992}. If a louver plate is introduced in the fluidized bed, the decrease in average bubble diameter is caused by bubble splitting events and the prevention of bubble coalescence \cite{Zhang.2009}. Additionally, louver plates reduce the gas and solid back-mixing in fluidized beds \cite{Zhang.2008}. Baffles enhance the uniform lateral bubble distribution \cite{Zhang.2020}.\\

Investigating the hydrodynamics within three-dimensional (3D) gas fluidized beds is challenging due to the lack of optical access to the interior of the bed. Hence, many previous experimental works simplified the 3D bed to pseudo-2D fluidized beds and used a high-speed camera to observe the system \cite{Laverman.2008, Mahajan.2018}. These systems have the drawback of the usually narrow wall separation distance of a few centimeters altering the hydrodynamics of the bed \cite{Geldart.1970}. Another approach is to insert intrusive probes into the 3D system such as capacitance probes \cite{Werther.1973} or pressure sensors \cite{Wei.2018}. The data from capacitance probes enable to conclude the average local bubble size, but the method is intrusive and thereby the hydrodynamics of the bed is disturbed by the probes \cite{Rudisuli.op.2013}. Tomographic techniques, such as positron emission particle tracking (PEPT) \cite{Parker.2008, Leadbeater.2022}, electric capacitance tomography \cite{Makkawi.2004}, gamma ray \cite{Mandal.2012}, X-ray tomography \cite{Grohse.1955, Mudde.2010}, or magnetic resonance imaging (MRI) \cite{Muller.2008} overcome both of these limitations and are hence promising experimental techniques to study the hydrodynamics of fluidized beds. A variety of other fast tomographic imaging methods for industrial process control have been recently developed and are described in this review \cite{Hampel2022}.\\

MRI enables to image opaque three-dimensional fluidized beds. Measurement slices can be oriented arbitrarily within the field of view and the method can image a variety of system properties beyond the spatial distribution of phases within the sample volume. MRI enables the direct measurement of particle velocities using phase contrast velocimetry with biploar motion-encoding gradient pulses \cite{Fukushima.AnnuRev1999}. Additionally, very recently, MRI has been demonstrated to be able to work as a three-dimensional thermometer in packed beds \cite{Serial.2023}. Yet, MRI was limited by its relatively low temporal resolution until a few years ago, which was recently improved considerably by Penn et al., using a combination of scan acceleration techniques, tailored receive coil arrays and engineered MRI-active particles \cite{Penn.2017}. Employing these techniques, it is possible to image local particle concentration and particle velocity in bubbling fluidized beds at high spatial and temporal resolution. In the past few years, real-time MRI was used to characterize the fluidization hydrodynamics (i.e. the number of bubbles, the bubble diameter, the bed height, and the particle velocities) in dry bubbling fluidized beds \cite{Penn.2018} and in bubbling fluidized beds into which small amounts of liquids were injected \cite{Boyce.AIChEJ2018}. The technique was also applied to map different regimes of jetting and bubbling in fluidized bed and to quantify the length of gas jets produced by injection ports \cite{Penn.CEJ2020}, and to study the effect of a horizontal tube on the hydrodynamics in fluidized beds \cite{Penn.2019}. The latter study found, that directly above a horizontal tube in a fluidized bed, one can find a significantly reduced bubble occurrence and particle velocities.\\

In this work, we used real-time MRI to investigate how different baffles within a three-dimensional fluidized bed affect the fluidization hydrodynamics of the bed. We compared three common baffle geometries and quantified the number of bubbles, the bubble sizes, the bubble shapes, and the particle velocities and compared the results to a fluidized bed without internal. In contrast to all previous works published using real-time MRI, here we also acquired and analyzed horizontal imaging slices in addition to vertical slices, allowing to study the horizontal distribution of bubbles at different vertical positions in the bed. 

\section{Methods}
\subsection{Fluidized bed} 
\begin{figure}[hbt!]
  \centering
  \includegraphics[width=8 cm]{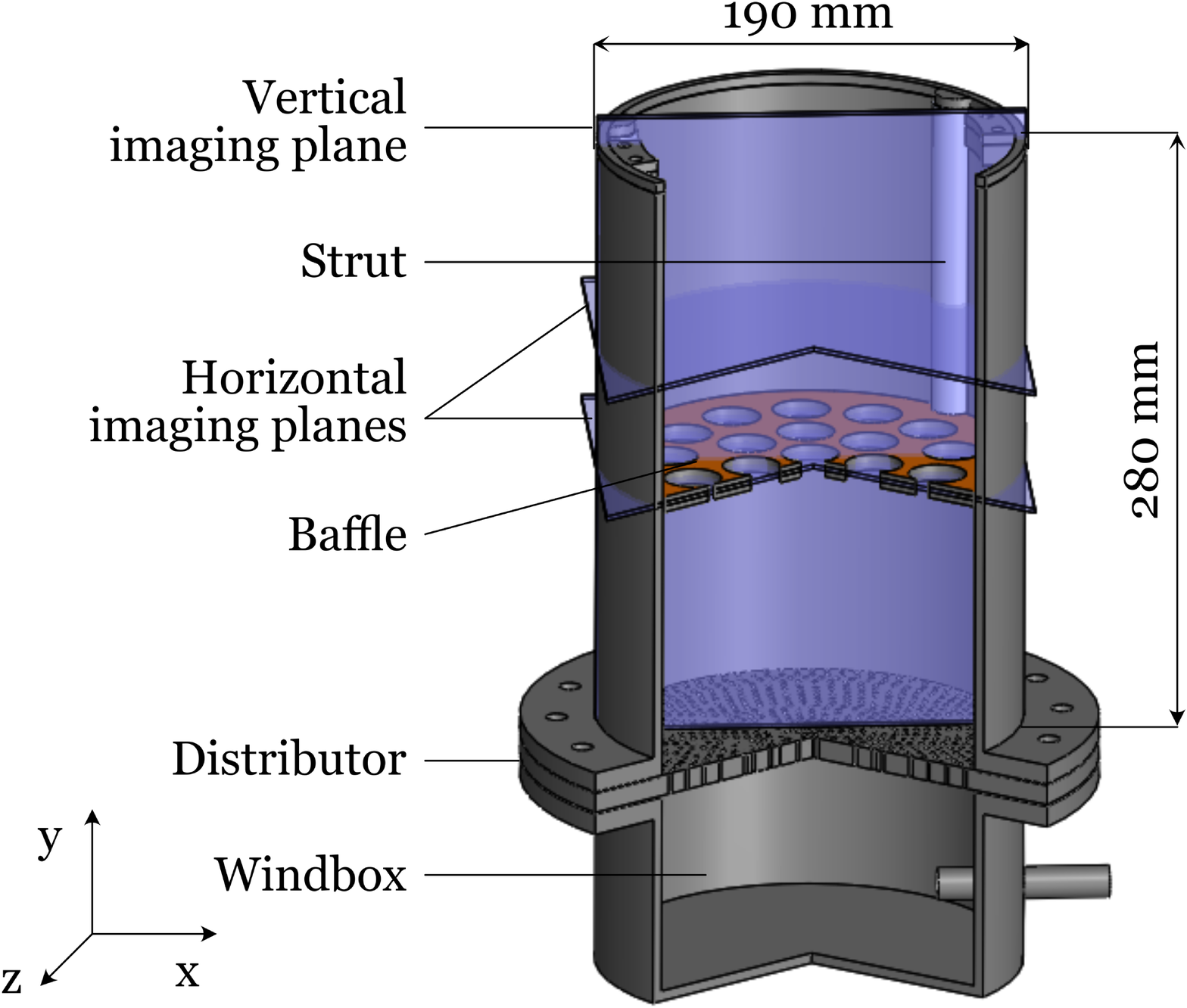}
         
         \caption{Sketch of the PMMA fluidized bed with a circled baffle included at a height of 120~mm above the distributor plate. The blue colored areas indicate the investigated imaging planes in MRI in horizontal direction at a height of 120~mm and 180~mm above the distributor plate and in vertical direction.}
         \label{fig:1}
\end{figure}

A cylindrical fluidized bed made from polymethylmethacrylat (PMMA) with an inner diameter of 190\,mm and a height of 280\,mm was used for all experiments of this study (Fig. ~\ref{fig:1}). The particle phase consisted of spherical agar particles filled with MRI-detectable medium-chain triglyceride (MCT) oil with a diameter of $1.02 \pm 0.12\,\mathrm{mm}$. The particles were filled in the fluidized bed up to a height of 210\,mm. The particle phase was fluidized with pressurized air using a mass flow controller (F-203AV, Bronkhorst High-Tech B.V.). The air entered the fluidized bed through a windbox of the same diameter as the bed and a height of 150 mm. The windbox and the fluidized bed were separated by a 10 mm thick PMMA distributor plate with 6414 evenly distributed holes of diameter of 0.5\,mm each. The minimum fluidization velocity was determined by measuring the pressure drop across the bed while defluidizing the bed using a pressure sensor connected to a data-acquisition interface (USB-6009 National Instruments, United States). More details about the  experimental setup without the baffles are described in \cite{Penn.2018}, and additional information on the particle properties and the MRI pulse sequences can be found in \cite{Penn.2017}.

\subsection{Baffle geometries} 
\begin{figure}[hbt!]
  \centering
  \includegraphics[width=8 cm]{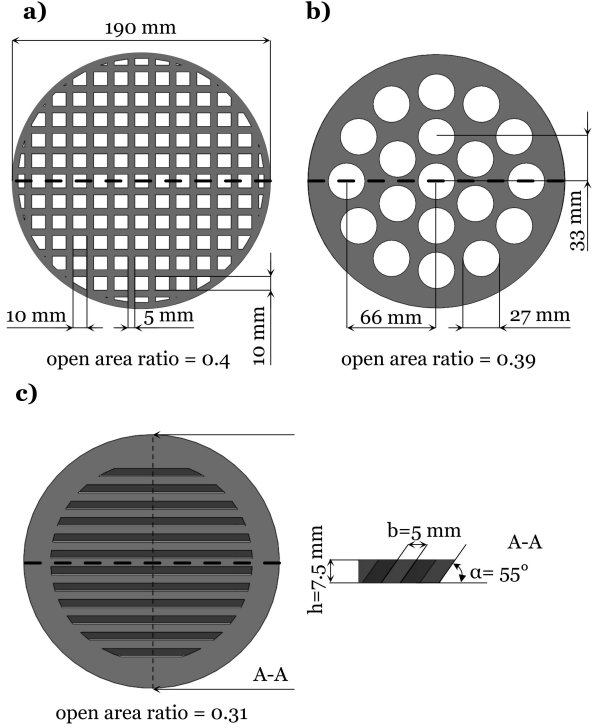}
     \caption{Sketch of the three baffle geometries with a thickness of 7.5~mm investigated in this study and compared to measurements in a bed without internal. The recorded imaging plane using MRI is located at the dashed horizontal line. (a) Grid structure with openings of a length of 10\,mm. (b) Circle structure with 19 circular openings. (c) Louver plate with a tilt angle of 55° and a web thickness of 5\,mm. The open area ratio for the corresponding baffle geometries is listed below the sketch.}
     \label{fig:2}
\end{figure}

Three different baffle geometries (Fig.~\ref{fig:2}) were studied: (a) a grid structure, (b) a circle structure, and (c) a louver plate. Each baffle was fixed in the fluidized bed column at a height of 120\,mm above the distributor plate using three struts connected to the upper edge of the fluidized bed (Fig. ~\ref{fig:1}). All baffle plates were 3D-printed from Polylactide on a Prusa i3 MK3S+ (Prusa Research, Czech Republic) with a plate thickness of 7.5\,mm. The baffle geometries were chosen based on the most commonly described designs in literature \cite{Dutta.1992,Zhang.2020,Yang.2003}. 

In Figure~\ref{fig:2}a, a baffle geometry based on a grid structure is illustrated. The grid structure comprised square openings with a length of 10\,mm and webs with a width of 5\,mm. The baffle geometry with circle structure illustrated in Figure~\ref{fig:2}b consisted of 19 circular openings with a diameter of 27\,mm. One opening was arranged in the center of the baffle and further openings were circumferentially arranged at two radial positions. The first radial position was 33\,mm and the second was 66\,mm. The louver plate illustrated in Figure~\ref{fig:2}c was constructed following existing guidelines \cite{Zhang.2009} described by Equation \ref{eq:louver_plate} with a tilt angle, $\alpha$, of 55°, a web thickness, $b$, of 5\,mm, and an height, $h$, of 7.5\,mm.

\begin{equation} 
b \approx  h \times \cos(\alpha)+(0-2)\,\mathrm{mm}
\label{eq:louver_plate}
\end{equation} 

The corresponding open area ratio of the three baffle geometries are collected in Figure~\ref{fig:2}. A fourth measurement series of the fluidized bed without internal was acquired for comparison. In the following, the CAD drawings of the baffles are used as symbols to distinguish between the different measurement series. Each setup has been measured with five different superficial gas velocities ($U_\mathrm{0}$) of 0, 1, 1.2, 1.5, and 2 times the minimum fluidization velocity ($U_\mathrm{mf}$).\\

\subsection{Magnetic resonance imaging} A clinical 3\,T MRI scanner (Philips Achieva 3.0 T, Philips Healthcare, The Netherlands) was used for the MRI measurements of this study. MRI measurements of local particle concentration and particle velocity were acquired in both a central vertical slice through the bed and two horizontal slices. The position of the vertical imaging plane is shown in Figure~\ref{fig:2} with a dashed line. The horizontal imaging planes were recorded 120\,mm and 180\,mm above the distributor plate. Both of these measurement types are based on time-efficient single-shot echo planar imaging (EPI) pulse sequences \cite{Stehling.1991} combined with parallel imaging \cite{Pruessmann1999} using a self-built 16-channel RF receiver array \cite{Penn.2017}. The particle velocity was directly be measured using phase contrast magnetic resonance velocimetry. Each measured voxel gives the average velocity of up to about 100 (in the unfluidized state) particles in that voxel. Employing this technique, the temporally averaged particle velocity and the velocity direction, the temporal fluctuation in particle velocity and the acceleration extracted from consecutive measurements can be measured. The experimental parameters and spatial and temporal resolution of the imaging sequences are listed in Table~\ref{tab:settings_MRI}. \\

\begin{table*}[hbt!]
    \centering
    \begin{tabular}{|p{3,7 cm}|c|c|c|c|}\hline
        &\multicolumn{2}{c|}{ \textbf{Local particle concentration}} & \multicolumn{2}{c|}{ \textbf{Particle velocity}} \\
        & vertical & horizontal & vertical & horizontal\\\hline
        \textbf{Number of frames per record} & 500 & 250 & 200 & 100 \\\hline
        \textbf{Field of View} & $200\times 300\,\mathrm{mm}$ & $200\times 200\,\mathrm{mm}$ & $200\times 300\,\mathrm{mm}$ & $200\times 200\,\mathrm{mm}$\\
        & ($x\,\times\,y$-direction) & ($x\,\times\,z$-direction) & ($x\,\times\,y$-direction)& ($x\,\times\,z$-direction)\\\hline 
        \textbf{Temporal resolution} & \multicolumn{2}{c|}{7\,ms} & \multicolumn{2}{c|}{21\,ms}\\\hline
        \textbf{Spatial resolution} & \multicolumn{2}{c|}{$3 \times 3 \times 10\,\mathrm{mm}$} &
    \multicolumn{2}{c|}{$3 \times 5 \times 15\,\mathrm{mm}$} \\\hline  
        \textbf{Flip angle (xy)} & \multicolumn{4}{c|}{15} \\\hline 
        \textbf{Repetition time $\mathbf{t_\mathrm{R}}$} &8.2\,ms& 8.2\,ms&1.9\,ms& 2.3\,ms\\\hline
        \textbf{Echo time $\mathbf{t_\mathrm{E}}$} &1.6\,ms& 2.6\,ms& 0.9\,ms& 1.3\,ms\\\hline
        \textbf{Velocity encoding} & - & - & 150\,cm/s in & 150\,cm/s \\
        &&& $x$ and $y$-direction & in $y$-direction\\\hline
        \textbf{Superficial gas velocity ($U_\mathrm{0}/U_\mathrm{mf}$)} &  \multicolumn{4}{c|}{0, 1, 1.2, 1.5, 2} \\\hline
        
        \end{tabular}
    \caption{Experimental parameters for the local particle concentration and particle velocity measurements.}
    \label{tab:settings_MRI}
\end{table*}

\subsection{Data processing} In all frames, the gas and particle phase were automatically segmented on the signal intensity of the spin density measurements. A spatially-dependent threshold was calculated based on a signal intensity image of the static bed without gas flow. 
Pixels below 75\,\% of the spatially-dependent average signal intensity were assigned to the gas phase. This approach hence corrects for the spatially varying sensitivity of the MRI detector elements. To avoid that pixel noise is interpreted as gas bubbles, bubbles with an area of less than $A_\mathrm{bub} < 2\pi\Delta x \Delta y$, where $\Delta x$ and $\Delta y$ are the resolution of particles in $x$ and $y$ direction, were assigned the particle phase. Bubbles that were erupting through the free surface of the bed, were excluded from the analysis as soon as they were connected with the ambient air above the bed. The equivalent bubble diameter, $d_\mathrm{bub}$, (Eq.~\ref{eq:equivalent_bubble_diameter}) was calculated and used for the subsequent quantitative analyses. 

\begin{equation}
    d_\mathrm{bub} = \sqrt{\frac{4A_\mathrm{bub}}{\pi}}    \label{eq:equivalent_bubble_diameter}
\end{equation}

Bed expansion was measured using the local particle concentration measurements. The average bed height was determined by averaging the highest particle of the bulk phase in each column. A radial weighting correction was introduced to correct for disproportionate overrepresentation of central regions in the 2D images compared to the 3D fluidized bed (Eq.~\ref{eq:filling height}). In Equation~\ref{eq:filling height} $h$ is defined as the expansion height of the fluidized bed, $r$ as the distance of a column to the center of the bed, and $M$, as the total column number.

\begin{equation}
    h = \frac{\sum\nolimits_{x=1}^M(h_i\frac{r_i}{r})}{\sum\nolimits_{i=1}^M\frac{r_i}{r}}    
    \label{eq:filling height}
\end{equation}

The particle velocity data were used to determine the average in-plane particle velocity, $|v_{xy}|$, (Eq.~\ref{eq:particle_velocity}), its standard deviation, $\mathrm{\sigma}(|v_{xy}|)$, (Eq.~\ref{eq:standard_deviation}) as well as the average acceleration, $|a_{xy}|$, (Eq.~\ref{eq:particle_acceleration}) of the particles (Fig.~\ref{fig:9}b-c, \ref{fig:10}, and S5-7). 
\begin{equation}
    |v_{xy}| = \frac{1}{N}\sum\nolimits_{i=1}^N{\sqrt{v_{x_i}^{2} + v_{y_i}^{2}}}
    \label{eq:particle_velocity}
\end{equation}

\begin{equation}
    \sigma(|v_{xy}|) = \sqrt{\frac{1}{N-1}\sum\nolimits_{i=1}^N{(v_{x_i}^{2} + v_{y_i}^{2})-|v_{xy}|}}
    \label{eq:standard_deviation}
\end{equation}

\begin{equation}
   |a_{xy}| = \frac{\sum\nolimits_{i=1}^{N-1}\sqrt{(v_{{x}_{i+1}}-v_{{x}_{i}})^2+ (v_{{y}_{i+1}}-v_{{y}_{i}})^2}}{\Delta t (N-1)}
    \label{eq:particle_acceleration}
\end{equation}

Where $N$ is the number of conducted measurements, $v_{x}$ the measured velocity in $x$ and $v_{y}$ in $y$-direction, and $\Delta t$ is the passed time between two consecutive measurements. Additionally, vector plots of averaged particle velocity were plotted in Figure~\ref{fig:9}a. For clarity of the plot, the data points were binned with a bin size of $4 \times 4$ pixels.

\section{Results and discussion}
\subsection{Spatial bubble distribution} 
\begin{figure*}[hbt!]
  \centering
  \includegraphics[width=14 cm]{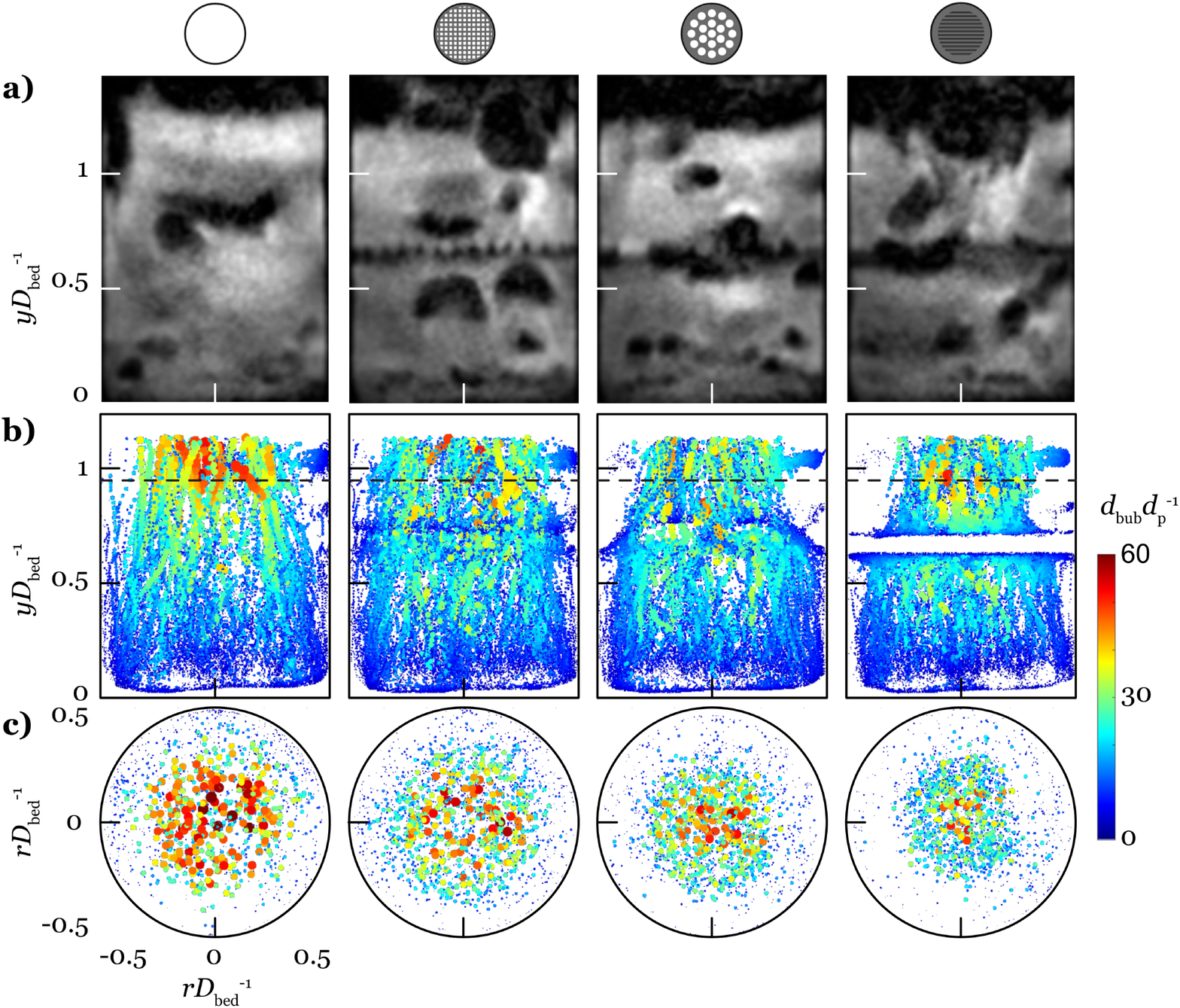}
         \caption{Local particle concentration and bubble size distribution for different baffle geometries at 1.5\,$U_\mathrm{mf}$. From left column to right column: bed without internal, grid structure, circle structure, and louver plate. (a) Exemplary MRI images of local particle concentration from a single frame out of the image time-series. (b) Spatial distribution of gas bubbles of a whole time-series of 1500 and 600 frames, respectively, in a central vertical slice through the bed and (c) a horizontal slice at a distance of 180\,mm (see dashed line in (b)) above the distributor plate. The bubble size is coded in color and plotted at their respective center of mass location. Respective figures for \,$U_\mathrm{0}$ = 1.2 and 2\,$U_\mathrm{mf}$ can be found in the supplementary information (Fig.~S1 and S2).}
         \label{fig:3}
\end{figure*}

Local particle concentration measurements (Fig.~\ref{fig:3}a) were used to study the position, size, and shape of gas bubbles in fluidized beds. Figures~\ref{fig:3}b and~\ref{fig:3}c show the spatial distribution of gas bubbles in the central vertical slice through the bed (Fig.~\ref{fig:3}b) and in a horizontal slice at a height of 180 mm above the distributor plate (Fig.~\ref{fig:3}c). Each position of each colored circle corresponds to the center of gravity of one individual gas bubble detected in the whole time-series of the acquisition. The size and color of the circles correspond to the size of the gas bubbles. The measurements of the four bed geometries for a superficial gas velocity of $1.5\,U_\mathrm{mf}$ were ordered from left to right as follows: bed without internal, grid structure, circle structure, and louver plate. In all investigated bed geometries, the bubble density is highest in the center and reduces towards the lateral edges of the fluidized bed. The spatial bubble distribution is considerably altered by introducing a baffle with circle structure or louver plate. The vertical bubble trajectories at the edge of the bed with circle structure and louver plate are angled towards the middle of the fluidized bed. This finding can be explained by the design of these two structures. Both baffles were constructed with solid plastic rings around the edges, 15\,mm for the circle structure and 20\,mm for the louver plate, preventing air from passing through. Air from this region, therefore, accelerates and flows towards the center of the bed. The trajectories of the rising bubbles of the bed without internal and grid structure are not exclusively directed to the middle of the bed. Here, no general acceleration of bubbles toward the middle of the fluidized bed can be detected. Compared to the non-directed trajectories in the bed without internal and the grid structure, the angled trajectories above the circle structure and the louver plate result in a decreasing homogeneity in the radial bubble distribution which can be observed both in the horizontal and vertical slices. Including a grid structure causes only a slight reduction of the bubble homogeneity in the radial direction. The number of bubbles at the edges of the bed with the grid structure and the bed without internal is similar. Whereas, at the center of the bed the grid structure increases the number of bubbles.\\

\subsection{Bubble number and size}

\begin{figure*}[hbt!]
  \centering
  \includegraphics[width=14 cm]{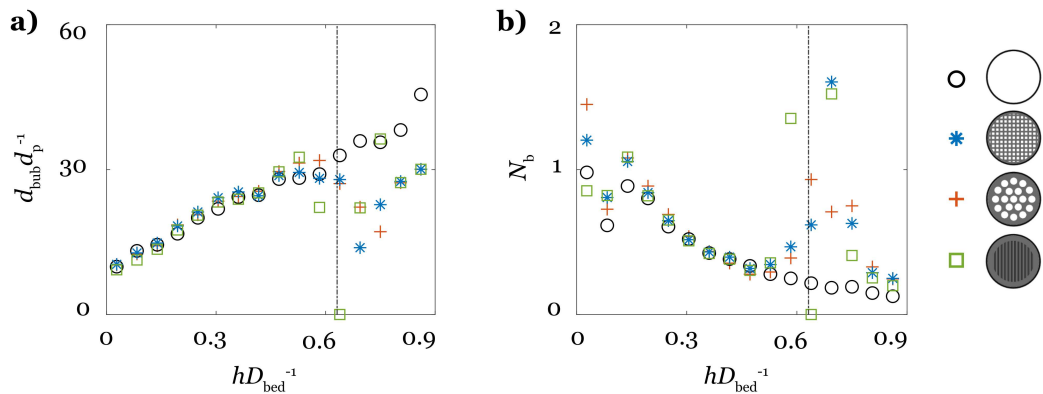}
         \caption{Average (a) equivalent bubble diameter and (b) number of bubble over bed height for a superficial gas velocity of 1.5\,$U_\mathrm{mf}$. The position of the baffle is depicted with a dashed line, if applicable. Figures for the cases of 1.2 and 2\,$U_\mathrm{mf}$ are shown in the supplementary material (Fig.~S3).}
\label{fig:4}
\end{figure*}
The equivalent bubble diameter in the bed without internal increases with vertical distance from the distributor plate as the color gradient across the bed height from predominantly yellow to red at the top shows (Fig.~\ref{fig:3}b left). For all investigated baffles in this study (Fig.~\ref{fig:3}b), we find a significantly reduced maximum bubble size compared to the bed without internal. Here, the dominant bubble color is yellow and green at the top of the fluidized bed. This observation indicates that the baffles limit the growth of gas bubbles in fluidized beds. For all setups, we observed a bubble accumulation in the upper-right corner of the bed. This bubble accumulation is not connected to the main field of bubbles by trajectories and corresponds to an area of low signal-to noise ratio in the images. A visual analysis of the data suggests that in this area particles are falsely assigned to the gas phase. To prevent this image artifact from affecting further analyses of the data, the edges of the fluidized bed were excluded in the following and only the middle part with a width of 120\,mm was considered. From this data, the average equivalent bubble diameter (Fig.~\ref{fig:4}a) and the average number of bubbles per frame and unit area (Fig.~\ref{fig:4}b) were calculated. The results for the bed without internal are in good agreement with previous data from \cite{Penn.2019} which in contrast to our study did not yet use a dynamic threshold. The bubble size increases monotonically with bed height. By introducing a baffle, the increase in bubble diameter is interrupted. At a superficial gas velocity of $1.5\,U_\mathrm{mf}$, the grid structure reduces the equivalent bubble diameter the most, down to 40\,\% compared to the bed without internal. The circle structure reduced to a minimum of 50\,\% and the louver plate to 60\,\%. Higher superficial gas velocities of 1.5 and 2\,$U_\mathrm{mf}$ result in a more pronounced reduction compared to 1.2\,$U_\mathrm{mf}$ (Fig.~S3). Directly downstream of the baffle, the average equivalent bubble diameter increases with the circle structure and the louver plate, but is not influenced by the grid structure. All beds with baffles show similar equivalent bubble diameters towards the top of the bed. Independent of the investigated superficial gas velocity, the equivalent diameter increases with distance above the baffle after reaching a local minimum. Besides the finding that baffles can reduce the average bubble diameter, it is important how long this effect can be observed at a distance from the baffle. This information can help to determine the construction distance between two baffles. The distance, at which the average equivalent bubble diameter equals that of the bed without internal, depends on the superficial gas velocity. At $1.2\,U_\mathrm{mf}$, the equivalent diameter returned to similar values as observed for the bed without internal at the maximum evaluated height for all investigated baffles. At 1.5 and 2\,$U_\mathrm{mf}$, the average bubble size in beds with baffles remains lower across the whole bed height. The average number of bubbles for the bed without internal decreases with increasing bed height, whereas the number of bubbles changes significantly in the vicinity of a baffle. Upstream of the baffle, the number of bubbles is elevated but decreases towards the value for the bed without internal at the top of the bed. 

The MR image time-series of the particle distribution shows how baffles decrease the equivalent bubble diameter, which can be observed in the videos provided in the supplementary  material. A strong interaction between gas bubbles and baffles was observable in the data. The baffles caused different bubble phenomena, including bubble splitting and the formation of air cushions. The appearance of these bubble phenomena has been reported previously in measurements using capacitance probes \cite{Dutta.1992} and optical imaging of  pseudo-2D fluidized beds \cite{Zhang.2009}. With our non-intrusive MRI experiments, we were able to confirm and visualize the occurrence of these bubble phenomena for 3D systems. Frequent bubble splitting was detected at the lower edge of all baffle geometries for all superficial gas velocities. 
 \begin{figure}[hbt!]
\centering
  \includegraphics[width=8 cm]{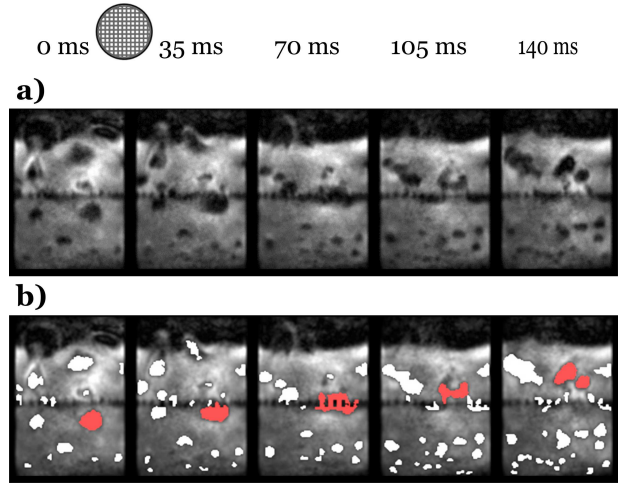}
\caption{Bubble splitting captured in six frames for the fluidized bed with an internal grid structure at 1.2\,$U_\mathrm{mf}$ with (a) MR images and (b) MR images overlaid by segmented images. The splitting bubble is marked in red colour.}
\label{fig:5}
\end{figure} 
 Each splitting event from one bubble into two smaller daughter bubbles decreases the average bubble diameter and increases the number of bubbles. It is thus assumed, that bubble splitting has an essential impact on the average bubble size and bubble distribution in systems with baffles. In Figure~\ref{fig:5}, one exemplary event of bubble splitting is illustrated (a) in MR images and (b) in MR images overlaid by the detected gas bubbles, with one bubble of interest being marked in red. The measurements of the setup with the grid structure for a superficial gas velocity of 1.2\,$U_\mathrm{mf}$ are illustrated with five frames of the image-time series. At times 35-105\,ms the bubble passes through the grid structure and subsequently splits at 105-140\,ms into two smaller daughter bubbles hence resulting in a reduced average bubble size and an increased total number of bubbles.

\subsection{Bed expansion}

\begin{figure}[hbt!]
\centering
 \includegraphics[width=8 cm]{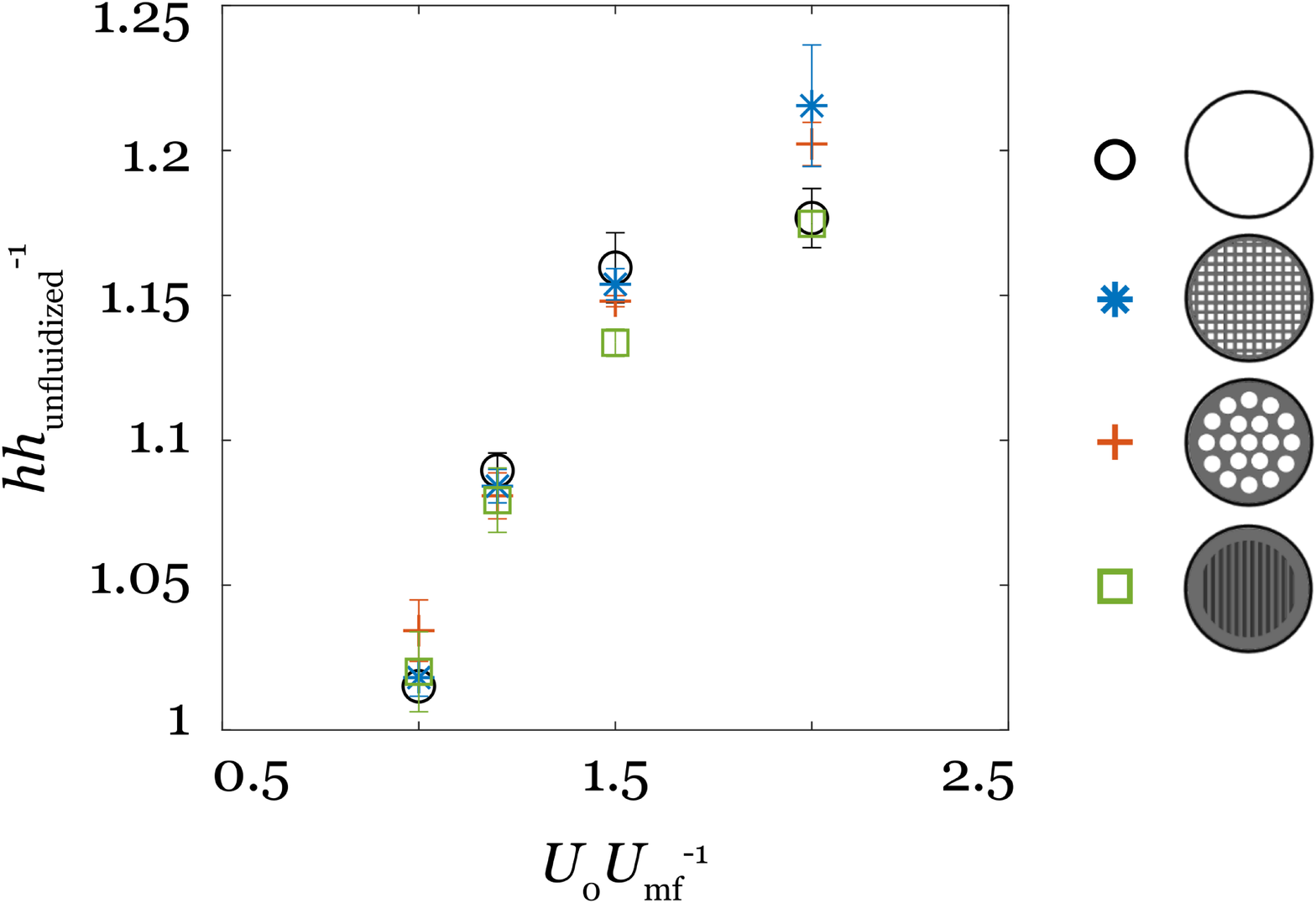}
  \caption{Expansion of the bed height with  superficial gas velocity for the bed without internal and the three different baffle types, normalized by the corresponding height of the unfluidized beds. Error bars were calculated from three experiments.}
  \label{fig:6}
\end{figure}

Figure~\ref{fig:6} shows the expanded bed height for different superficial gas velocities. The bed expansion provides information about the total gas holdup in the bed. In all geometries, the expansion increases with increasing gas volume flow. 

\begin{figure}[hbt!]
  \centering
  \includegraphics[width=8cm]{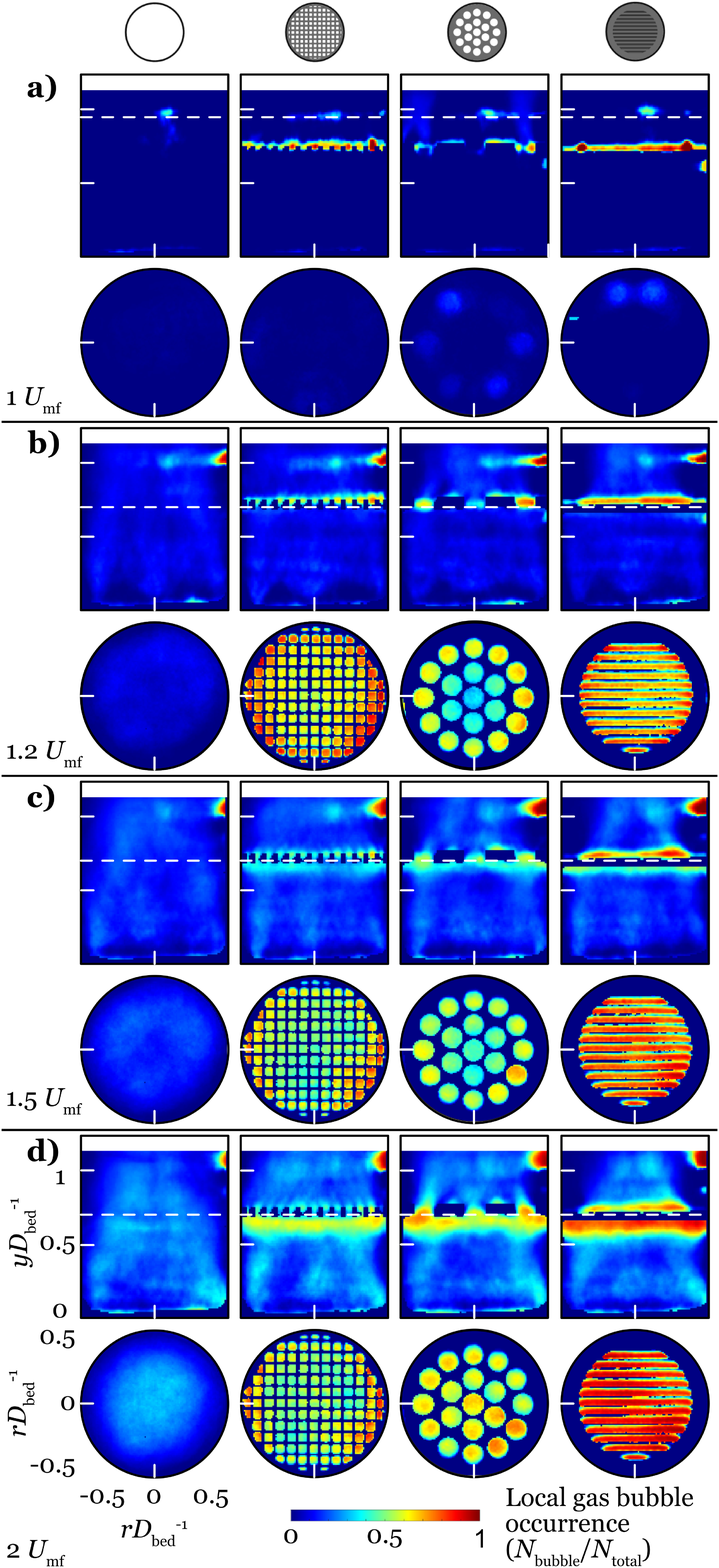}
         \caption{Heat maps of local gas bubble occurrence in the bed. Both horizontal and vertical slices are shown. Blue areas show regions with almost no bubbles, red areas show regions of almost exclusively bubbles. The four panels  correspond to the four different superficial gas velocities $U_\mathrm{0}$: (a) 1, (b) 1.2, (c) 1.5, and (d) 2\,$U_\mathrm{mf}$. For superficial gas velocities between 1.2 and 2\,$U_\mathrm{mf}$ the measurements show an artifact in the upper right corner (compare Figure\,\ref{fig:3}b).}
         \label{fig:8}
\end{figure}

For the superficial gas velocities below 1.5\,$U_\mathrm{mf}$, the expansion is unchanged by the presence of baffles. For the highest investigated superficial gas velocity of 2\,$U_\mathrm{mf}$, baffles influence the bed expansion. The bed without internal and the louver plate exhibit the lowest expansion of approximately 18\,\% compared to the unfluidized state. The circle and grid structures expand the bed by up to 21\,\%. There are two possible explanations for why introducing a baffle increases the expansion of the bed height: (a) Baffles increase the gas holdup inside the fluidized bed by increasing the residence time of the gas. The residence time can for example be increased by air cushions (Fig.~\ref{fig:8}) under the baffle structure. (b) The bubbles which on average have a larger diameter in the bed without internal, are earlier connected with the surrounding and thus lower the average height of the highest pixel per column connected to the bulk phase.\\

\subsection{Bubble shape}
\begin{figure}[hbt!]
  \centering
  \includegraphics[width=8cm]{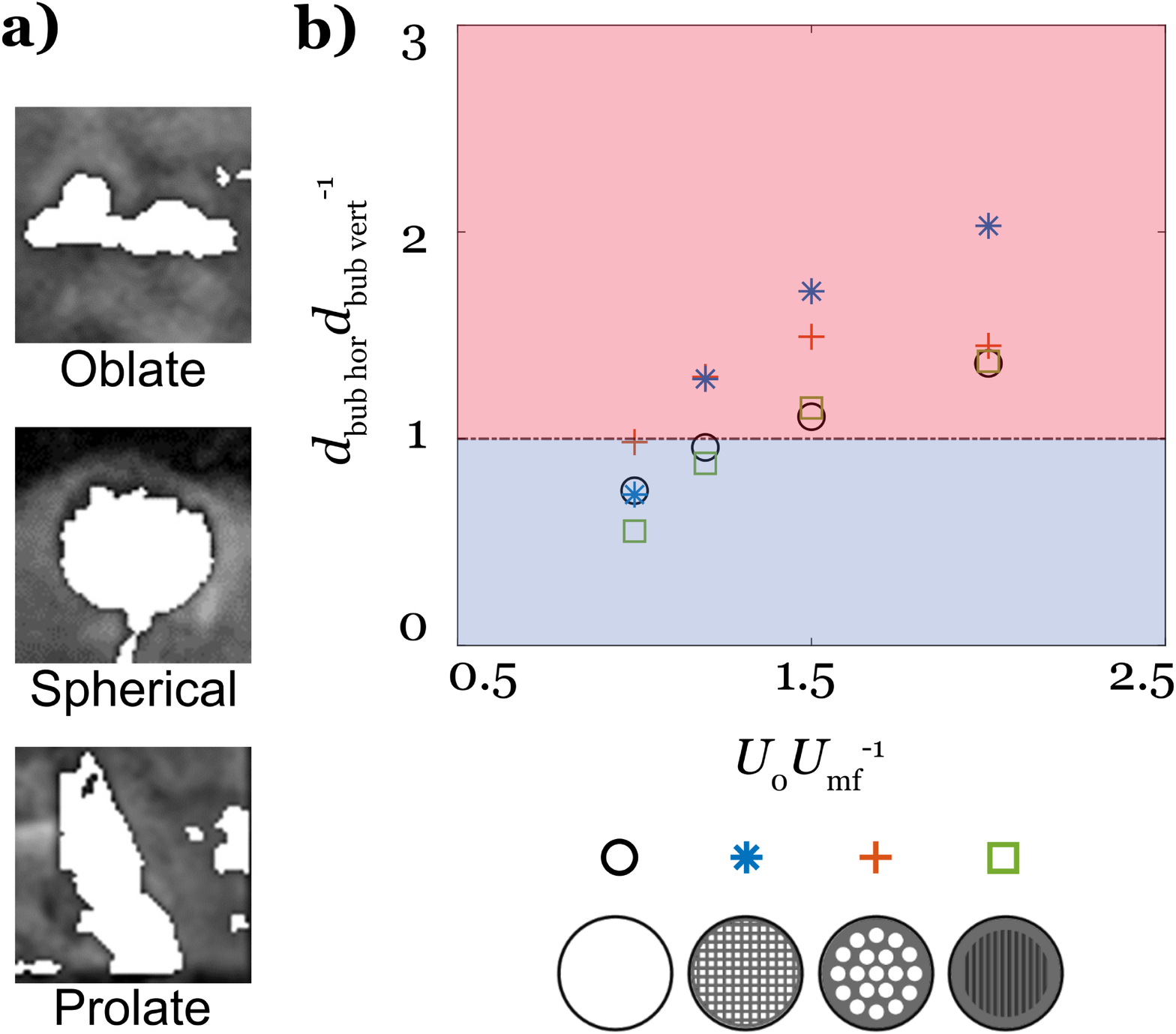}
   \caption{(a) Exemplary images of bubbles shaped as a prolate, sphere, and oblate. (b) Ratio of the average diameter in horizontal divided by vertical orientation over velocity ratio. If the diameter ratio is below one, bubbles are shaped as a prolates, for a ratio of one as spheres and above one as oblates.}
    \label{fig:7}
\end{figure}

The shape of gas bubbles affects their surface-to-volume ratio and is therefore of interest when characterizing bubbling fluidized beds. Here we used a simple ratio of average equivalent bubble diameter in horizontal vs. vertical orientation to analyze the influence of baffles on the shape of detected gas bubbles in fluidized beds. If the ratio equals one, the bubbles are spherical. For values $\neq$1, the bubbles are ellipsoidally shaped.  If the ratio is $<$1 the bubbles are prolates, $>$1, they exhibit an oblate shape (Fig~\ref{fig:7}a). More detailed information about the shape of droplets in multiphase flow can be found in \cite{Clift.2013}. For all beds studied, the diameter ratio increases with superficial gas velocity, and hence the bubbles become flatter. The bed with a louver plate showed a similar change of the diameter ratio as the bed without internal. The grid and circle structure increase the average diameter ratio compared to the bed without internal leading to more oblate (flatter) bubbles. For the grid structure, the difference compared to the bed without internal increases with increasing superficial gas velocity. Whereas, for the circle structure the diameter ratio is higher compared to the bed without internal for gas velocities between 1 and 1.5\,$U_\mathrm{mf}$. For a gas velocity of 2\,$U_\mathrm{mf}$, the bubble shape on average is similar to those in a bed without internal. In literature, the bubble aspect ratio, which is the ratio of the bubble width divided by the bubble height in vertical orientation, and the shape factor, which is the equivalent bubble diameter times $\pi$ divided by the bubble circumference, are used to describe the bubble shape in fluidized beds. Compared to the ratio we used to evaluate the bubble shape, aspect ratio and shape factor can be determined by measurements in vertical orientation alone. Bubbles are divided in groups of elongated ellipsoids (aspect ratio $<$ 1), spheres (aspect ratio = 1), and spherical caps (aspect ratio $>$ 1). In a pseudo-2D fluidized bed, Caicedo et al. \cite{Caicedo.2003} investigated the shape factor and the aspect ratio with digital image analysis. The shape factor and aspect ratio were normally distributed for constant experiments settings. Neither shape factor nor aspect ratio were strongly influenced by the particle diameter, column width, bed height, and relative humidity. Only the superficial gas velocity influenced the shape factor. Boyce et al. \cite{Boyce.2019} used MRI to determine the aspect ratio of single injected bubbles in three-dimensional fluidized beds with a superficial gas velocity of 1\,$U_\mathrm{mf}$. The aspect ratio decreases with increasing bed height until the bubbles pinch-off from the injection orifice. Up from the bubble pinch-off the aspect ratio increases with increasing bed height. Higher injection times and thereby bubble volume lead to smaller aspect ratios in average.\\

\subsection{Bubble phenomena}
By mapping the average number of detected bubbles per frame, information on bubble phenomena like air cushions and gas channeling can be extracted. Figure~\ref{fig:8} shows the ratio of a pixel detected as a gas bubble, $N_\mathrm{gas}$, divided by the total number of measurements per pixel, $N_\mathrm{total}$, in vertical orientation (upper row). The dashed line provides information on whether the horizontal orientated maps shown below are recorded at a distance of 120 or 180\,mm above the distributor. If $N_\mathrm{gas}/N_\mathrm{total}$ of a pixel equals one, a bubble was detected in all evaluated frames. The heat maps are ordered in rows related to the superficial gas velocity, which is (a) 1\,$U_\mathrm{mf}$ (b) 1.2\,$U_\mathrm{mf}$ (c) 1.5\,$U_\mathrm{mf}$, and (d) 2\,$U_\mathrm{mf}$. In the vertical orientation, a high bubble occurrence is detected below the baffle structure, indicating that air cushions form under the baffle. The occurrence of those air cushions under louver plates was already described for pseudo-2D fluidized beds in literature \cite{Zhang.2009, Yang.2003}. Zhang et al. showed that with increasing superficial gas velocity the height of the gas cushion increases. For our measurements, the thickness and incidence of these cushions are dependent on the baffle geometry and increase with higher superficial gas velocities, as well. Especially the louver plate causes relatively thick and permanent air cushions at a volume flow of 2\,$U_\mathrm{mf}$. The detected cushions indicate that the particles do not change their position from below to above the baffle and vice versa, effectively causing a segmentation of the bed into two compartments. This trend can be seen even more clearly in the horizontally orientated heat maps (Fig.~\ref{fig:8}b-d, lower row). At 120\,mm above the distributor plate, which is located at the height of the baffle, in all baffles, 0.5 or more bubbles per frame are detected for superficial gas velocities between 1.2 and 2\,$U_\mathrm{mf}$. The air bubbles disturb the particle exchange. This restriction in mass transfer can reduce the efficiency of the fluidized bed in certain applications. Additionally, the formation of bubbles above the baffle, which can be detected for the grid structure and the louver plate, can be caused by locally higher gas velocities. The baffles reduce the area through which gas flows and thus increase the average velocity of the gas.\\

Channeling lowers the efficiency of a process by decreasing the residence time of the gas and the contact area between gas and particles \cite{Briens.1997}. In the following, it is evaluated if introducing a baffle causes channeling for superficial gas velocity of $U_\mathrm{0}$ =\,$U_\mathrm{mf}$. The circle structure increases bubble formation above two circled openings in the vertical imaging plane (Fig.~\ref{fig:8}a). In horizontal orientation, the bubble frequency in the bed without internal and in the setup with the grid structure is low and non-directional. In contrast, the number of bubbles for the circle structure and the louver plate is increased. The location of the bubbles is over the outer circled openings for the circle structure and over one of the outermost openings for the louver plate. Therefore, it can be concluded that the louver plate and the circle structure cause channeling for superficial gas velocities of the minimum fluidization velocity. Whereas for the grid structure and the bed without internal no channeling can be detected.\\

\subsection{Particle velocity and acceleration} 
\begin{figure*}[hbt!]
  \centering
  \includegraphics[width=14 cm]{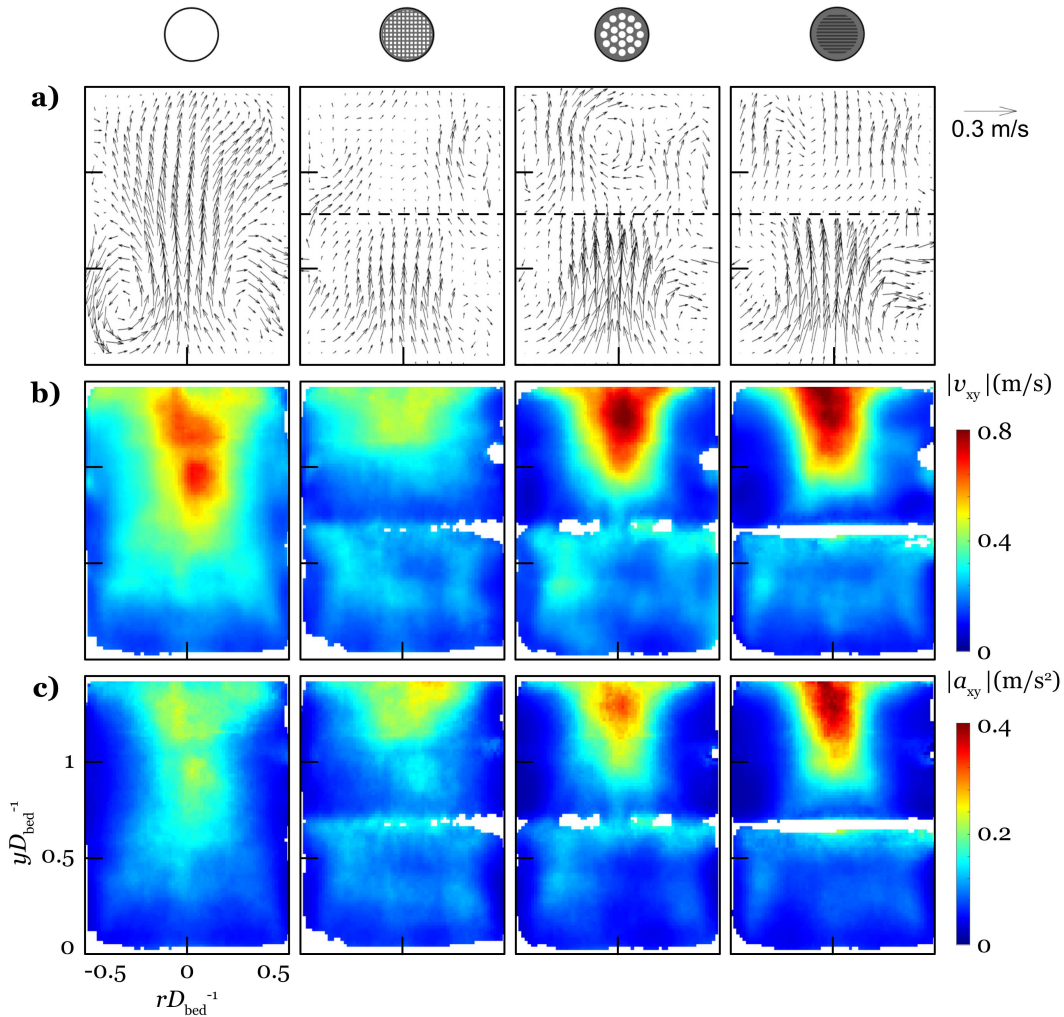}
         \caption{Particle velocity: (a)~vector field of the temporally particle velocity (b)~the temporally averaged particle velocity, (c)~the particle acceleration for measurements with a superficial gas velocity of 2\,$U_\mathrm{mf}$.}
         \label{fig:9}
\end{figure*}

Figure~\ref{fig:9}a shows vector plots of the temporally averaged particle velocity in the different bed types at a gas velocity of 2\,$U_\mathrm{mf}$. The corresponding plots for gas velocities of 0, 1.2, and 1.5\,$U_\mathrm{mf}$ are shown in the supplementary material (Fig.~S4). The bed without internal exhibits the typical convective flow behavior with particles rising in the center of bed. This velocity pattern has been described previously \cite{Li.2014}. The presence of a baffle alters this convective pattern: All the baffles studied in this work showed more complex patterns of average particle velocity and flow direction. For the grid structure, the sum of the velocities above the structure is lower than for the other setups. The circle structure causes vortex buildup in the upper part of the setup.

Figure~\ref{fig:9}b-c shows the temporally averaged particle velocity and the particle acceleration (derived mathematically from the velocity measurements, Eq.~\ref{eq:particle_acceleration}). The standard deviation for superficial gas velocities between 1.2 and 2\,$U_\mathrm{mf}$ can be found in the supplementary material (Fig.~S5). The central region close to the free surface of the beds exhibits the highest temporally averaged particle velocity. This finding is in line with previous works of freely bubbling fluidized beds and beds with horizontal tubes \cite{Penn.2019}. Towards the edges of the bed, the particle velocity decreases due to wall effects for all investigated superficial gas velocities. This effect is most pronounced in the beds with the circle structure and the louver plate. One reason for the low velocities at the edges of the fluidized bed could be caused by the design of the rather thick ring around the edge of the baffles, which might direct air toward the center of the bed. The grid structure reduces averaged particle velocity in the bed. All baffles decrease the velocity significantly in the lower two-thirds of the bed compared to the bed without internal. In total, the absolute average particle velocity decreases by introducing a baffle. The spatial distribution of the maximum velocities (Fig.~\ref{fig:9}b) seems to contradict the vector fields (Fig.~\ref{fig:9}a), where the highest velocities show up in the bottom half of the fluidized bed. This difference can be explained, since velocities are considered as non-directed for the temporally averaged particle field and directed for the vector fields. The rapid change in direction of the velocities can also be recorded in the mean accelerations between successive recordings. In all beds studied here, the acceleration is highest at the center top of the bed. Here the particles change their velocity the most. In beds with a baffle, the average acceleration increases slightly directly below the structure. All baffle geometries increase the maximum particle acceleration but reduce the average particle acceleration, hence exhibit a more heterogeneous hydrodynamic behavior. The higher maximum average acceleration could lead to increased particle abrasion and hence reduce overall process efficiency. The acceleration provides information about the temporal change of particle velocities between consecutive frames. Whereas, with the fluctuation in particle velocity the deviation to the average is calculated. The fluctuation in particle velocity is shown in Figure~S5c. The grid structure is characterized by low fluctuation in velocity. 

\begin{figure*}[hbt!]
  \centering
  \includegraphics[width=14 cm]{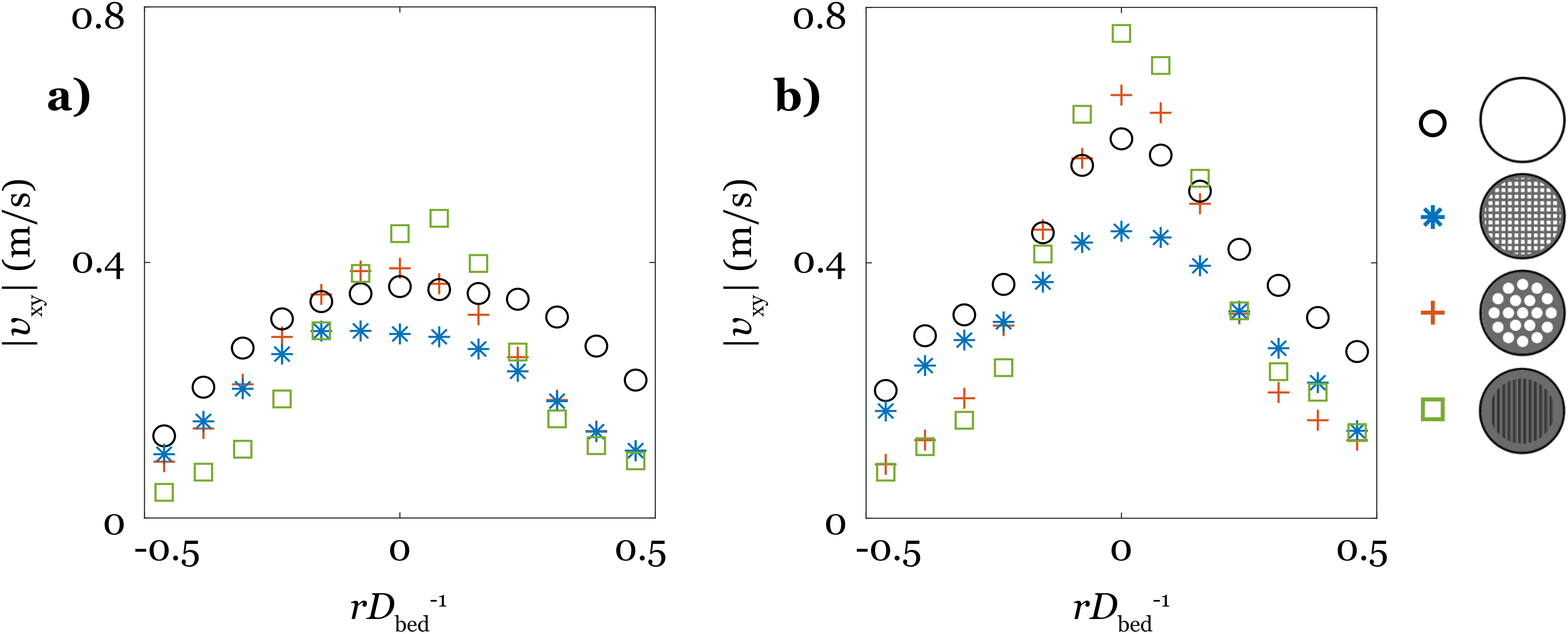}
         \caption{Temporally averaged particle velocity over bed radius at a height of 180\,mm above the distributor plate. For the evaluation vertical orientated measurements with (a) 1.5 and (b) 2\,$U_\mathrm{mf}$ were used.}
        \label{fig:10}
\end{figure*}

To compare the homogeneity of the temporally averaged particle velocity between the different setups, the velocities of the vertically orientated measurements are plotted over the radius for a distance of 180\,mm above the distributor plate (Fig.~\ref{fig:10}). The grid structure exhibits the most homogeneous velocity distribution. The bed without internal shows slightly  more heterogeneous distribution, especially at 2\,$U_\mathrm{mf}$. For the circle structure and louver plate a sharper maximum towards the middle is observed. Since for a local homogeneous mass transfer in the fluidized bed the particle velocities have to be as homogeneous as possible, the circle structure and louver plate have a negative impact on the process. 

\section{Conclusion}
In this work, we used real-time MRI to study the effect of different baffles on the fluidization hydrodynamics in a three-dimensional gas-solid fluidized bed model. All investigated baffles decreased the size and increased the number of bubbles in the fluidized bed. This effect was most pronounced upstream of the baffles and extended over a larger vertical distance for higher superficial gas velocities $U_\mathrm{0}$. Bubble phenomena, such as air cushions and bubble splitting, have been observed to form below the different baffle structures. This effect becomes more prominent with increasing $U_\mathrm{0}$. The grid and circle structure increased the average bed expansion for superficial gas velocities of 2\,$U_\mathrm{mf}$ and led to more oblate bubble shapes. The back-mixing of particles between areas above and below the baffle was found to be reduced by air flowing through the openings of the baffle and the formation of air cushions below the baffle. Already at the minimum fluidization velocity $U_\mathrm{mf}$, the circle structure and the louver plate showed gas channeling due to the locally increased gas velocity. All baffles increase the heterogeneity of particle convection. The circle structure and the louver plate additionally decreased the homogeneity of the particle velocity and the lateral bubble distribution. \\

Our findings can contribute to a more profound understanding of the effect of baffles inside a fluidized bed on the fluidization hydrodynamics. The findings could contribute to deciding if including a baffle is beneficial for a process. Additionally, the comparison between the three baffle types showed how internal design can affect in particular the homogeneity of the particle velocity and bubble distribution. The grid structure has proven to be advantageous in this respect and should be further investigated in the future.  Moreover, the effect of multiple baffles in a fluidized bed with a higher bed diameter and height could be evaluated.

\section*{Acknowledgement}
The authors thank Klaas P. Pruessmann from the University of Zurich and ETH Zurich for granting access to their MRI facility. The MR image data collected in this study will be made publicly available via the website of the institute. 


\bibliographystyle{elsarticle-num}
\bibliography{Literaturverzeichnis/cas-refs}

\begin{thebibliography}{10}
\expandafter\ifx\csname url\endcsname\relax
  \def\url#1{\texttt{#1}}\fi
\expandafter\ifx\csname urlprefix\endcsname\relax\def\urlprefix{URL }\fi
\expandafter\ifx\csname href\endcsname\relax
  \def\href#1#2{#2} \def\path#1{#1}\fi

\bibitem{Geldart.1973}
D.~Geldart, Types of gas fluidization, Powder Technology 7~(5) (1973) 285--292.
\newblock \href {https://doi.org/10.1016/0032-5910(73)80037-3}
  {\path{doi:10.1016/0032-5910(73)80037-3}}.

\bibitem{Cuenca.1995}
M.~A. Cuenca, E.~J. Anthony (Eds.), Pressurized Fluidized Bed Combustion,
  {Springer Netherlands}, Dordrecht, 1995.
\newblock \href {https://doi.org/10.1007/978-94-011-0617-7}
  {\path{doi:10.1007/978-94-011-0617-7}}.

\bibitem{Amiri.2016}
Z.~Amiri, S.~Movahedirad, M.~Shirvani, Particles mixing induced by bubbles in a
  gas-solid fluidized bed, AIChE Journal 62~(5) (2016) 1430--1438.
\newblock \href {https://doi.org/10.1002/aic.15150}
  {\path{doi:10.1002/aic.15150}}.

\bibitem{Dutta.1992}
S.~Dutta, G.~D. Suciu, An experimental study of the effectiveness of baffles
  and internals in breaking bubbles in fluid beds, JOURNAL OF CHEMICAL
  ENGINEERING OF JAPAN 25~(3) (1992) 345--348.
\newblock \href {https://doi.org/10.1252/jcej.25.345}
  {\path{doi:10.1252/jcej.25.345}}.

\bibitem{Zhang.2009}
Y.~Zhang, J.~R. Grace, X.~Bi, C.~Lu, M.~Shi, Effect of louver baffles on
  hydrodynamics and gas mixing in a fluidized bed of fcc particles, Chemical
  Engineering Science 64~(14) (2009) 3270--3281.
\newblock \href {https://doi.org/10.1016/j.ces.2009.04.017}
  {\path{doi:10.1016/j.ces.2009.04.017}}.

\bibitem{Zhang.2008}
Y.~Zhang, C.~Lu, J.~R. Grace, X.~Bi, M.~Shi, Gas back-mixing in a
  two-dimensional baffled turbulent fluidized bed, Industrial {\&} Engineering
  Chemistry Research 47~(21) (2008) 8484--8491.
\newblock \href {https://doi.org/10.1021/ie800906n}
  {\path{doi:10.1021/ie800906n}}.

\bibitem{Zhang.2020}
Y.~Zhang, Baffles and aids to fluidization, in: J.~Grace, X.~Bi, N.~Ellis
  (Eds.), Essentials of Fluidization Technology, Wiley, 2020, pp. 431--455.
\newblock \href {https://doi.org/10.1002/9783527699483.ch18}
  {\path{doi:10.1002/9783527699483.ch18}}.

\bibitem{Laverman.2008}
J.~A. Laverman, I.~Roghair, M.~S. {van Annaland}, H.~Kuipers, Investigation
  into the hydrodynamics of gas--solid fluidized beds using particle image
  velocimetry coupled with digital image analysis, The Canadian Journal of
  Chemical Engineering 86~(3) (2008) 523--535.
\newblock \href {https://doi.org/10.1002/cjce.20054}
  {\path{doi:10.1002/cjce.20054}}.

\bibitem{Mahajan.2018}
V.~V. Mahajan, J.~T. Padding, T.~M.~J. Nijssen, K.~A. Buist, J.~A.~M. Kuipers,
  Nonspherical particles in a pseudo-2d fluidized bed: Experimental study,
  AIChE Journal 64~(5) (2018) 1573--1590.
\newblock \href {https://doi.org/10.1002/aic.16078}
  {\path{doi:10.1002/aic.16078}}.

\bibitem{Geldart.1970}
D.~Geldart, The size and frequency of bubbles in two- and three-dimensional
  gas-fluidised beds, Powder Technology 4~(1) (1970) 41--55.
\newblock \href {https://doi.org/10.1016/0032-5910(70)80007-9}
  {\path{doi:10.1016/0032-5910(70)80007-9}}.

\bibitem{Werther.1973}
J.~Werther, O.~Molerus, The local structure of gas fluidized beds ---i. a
  statistically based measuring system, International Journal of Multiphase
  Flow 1~(1) (1973) 103--122.
\newblock \href {https://doi.org/10.1016/0301-9322(73)90007-4}
  {\path{doi:10.1016/0301-9322(73)90007-4}}.

\bibitem{Wei.2018}
L.~Wei, Y.~Lu, J.~Zhu, G.~Jiang, J.~Hu, H.~Teng, Effect of cohesive powders on
  pressure fluctuation characteristics of a binary gas-solid fluidized bed,
  Korean Journal of Chemical Engineering 35~(10) (2018) 2117--2126.
\newblock \href {https://doi.org/10.1007/s11814-018-0115-8}
  {\path{doi:10.1007/s11814-018-0115-8}}.

\bibitem{Rudisuli.op.2013}
M.~R{\"u}dis{\"u}li, T.~J. Schildhauer, S.~Biollaz, J.~R. {van Ommen}, 18 -
  measurement, monitoring and control of fluidized bed combustion and
  gasification, in: F.~Scala (Ed.), Fluidized-bed technologies for near-zero
  emission combustion and gasification, Woodhead publishing series in energy,
  {Woodhead publ}, Oxford, op. 2013, pp. 813--864.
\newblock \href {https://doi.org/10.1533/9780857098801.3.813}
  {\path{doi:10.1533/9780857098801.3.813}}.

\bibitem{Parker.2008}
D.~J. Parker, X.~Fan, Positron emission particle tracking---application and
  labelling techniques, Particuology 6~(1) (2008) 16--23.
\newblock \href {https://doi.org/10.1016/j.cpart.2007.10.004}
  {\path{doi:10.1016/j.cpart.2007.10.004}}.

\bibitem{Leadbeater.2022}
T.~W. Leadbeater, J.~P.~K. Seville, D.~J. Parker, On trajectory and velocity
  measurements in fluidized beds using positron emission particle tracking (
  pept ), The Canadian Journal of Chemical Engineering (2022).
\newblock \href {https://doi.org/10.1002/cjce.24622}
  {\path{doi:10.1002/cjce.24622}}.

\bibitem{Makkawi.2004}
Y.~T. Makkawi, P.~C. Wright, Electrical capacitance tomography for conventional
  fluidized bed measurements---remarks on the measuring technique, Powder
  Technology 148~(2-3) (2004) 142--157.
\newblock \href {https://doi.org/10.1016/j. powtec.2004.09.006}
  {\path{doi:10.1016/j. powtec.2004.09.006}}.

\bibitem{Mandal.2012}
D.~Mandal, V.~K. Sharma, H.~J. Pant, D.~Sathiyamoorthy, M.~Vinjamur, Quality of
  fluidization in gas--solid unary and packed fluidized beds: An experimental
  study using gamma ray transmission technique, Powder Technology 226 (2012)
  91--98.
\newblock \href {https://doi.org/10.1016/j.powtec.2012.04.022}
  {\path{doi:10.1016/j.powtec.2012.04.022}}.

\bibitem{Grohse.1955}
E.~W. Grohse, Analysis of gas-fluidized solid systems by x-ray absorption,
  AIChE Journal 1~(3) (1955) 358--365.
\newblock \href {https://doi.org/10.1002/aic.690010315}
  {\path{doi:10.1002/aic.690010315}}.

\bibitem{Mudde.2010}
R.~F. Mudde, Time-resolved x-ray tomography of a fluidized bed, Powder
  Technology 199~(1) (2010) 55--59.
\newblock \href {https://doi.org/10.1016/j. powtec.2009.04.021}
  {\path{doi:10.1016/j. powtec.2009.04.021}}.

\bibitem{Muller.2008}
C.~R. M{\"u}ller, D.~J. Holland, A.~J. Sederman, M.~D. Mantle, L.~F. Gladden,
  J.~F. Davidson, Magnetic resonance imaging of fluidized beds, Powder
  Technology 183~(1) (2008) 53--62.
\newblock \href {https://doi.org/10.1016/j.powtec.2007.11.029}
  {\path{doi:10.1016/j.powtec.2007.11.029}}.

\bibitem{Hampel2022}
U.~Hampel, L.~Babout, R.~Banasiak, E.~Schleicher, M.~Soleimani, T.~Wondrak,
  M.~Vauhkonen, T.~Lähivaara, C.~Tan, B.~Hoyle, A.~Penn, A review on fast
  tomographic imaging techniques and their potential application in industrial
  process control, Sensors 22~(6) (2022).
\newblock \href {https://doi.org/10.3390/s22062309}
  {\path{doi:10.3390/s22062309}}.

\bibitem{Fukushima.AnnuRev1999}
E.~Fukushima, Nuclear magnetic resonance as a tool to study flow, Annual Review
  of Fluid Mechanics 31~(1) (1999) 95--123.
\newblock \href {https://doi.org/10.1146/annurev.fluid.31.1.95}
  {\path{doi:10.1146/annurev.fluid.31.1.95}}.

\bibitem{Serial.2023}
M.~R. Serial, S.~Benders, P.~Rotzetter, D.~L. Brummerloh, J.~P. Metzger, S.~P.
  Gross, J.~Nussbaum, C.~R. Müller, K.~P. Pruessmann, A.~Penn, Temperature
  distribution in a gas-solid fixed bed probed by rapid magnetic resonance
  imaging, Chemical Engineering Science 269 (2023) 118457.
\newblock \href {https://doi.org/https://doi.org/10.1016/j .ces.2023.118457}
  {\path{doi:https://doi.org/10.1016/j .ces.2023.118457}}.

\bibitem{Penn.2017}
A.~Penn, T.~Tsuji, D.~O. Brunner, C.~M. Boyce, K.~P. Pruessmann, C.~R.
  M{\"u}ller, Real-time probing of granular dynamics with magnetic resonance,
  Science advances 3~(9) (2017) e1701879.
\newblock \href {https://doi.org/10.1126/sciadv.1701879}
  {\path{doi:10.1126/sciadv.1701879}}.

\bibitem{Penn.2018}
A.~Penn, C.~M. Boyce, T.~Kovar, T.~Tsuji, K.~P. Pruessmann, C.~R. M{\"u}ller,
  Real-time magnetic resonance imaging of bubble behavior and particle velocity
  in fluidized beds, Industrial {\&} Engineering Chemistry Research 57~(29)
  (2018) 9674--9682.
\newblock \href {https://doi.org/10.1021/acs.iecr.8b00932}
  {\path{doi:10.1021/acs.iecr.8b00932}}.

\bibitem{Boyce.AIChEJ2018}
C.~M. Boyce, A.~Penn, K.~P. Pruessmann, C.~R. Müller, Magnetic resonance
  imaging of gas–solid fluidization with liquid bridging, AIChE Journal
  64~(8) (2018) 2958--2971.
\newblock \href {https://doi.org/https://doi.org/10.1002/aic.16036}
  {\path{doi:https://doi.org/10.1002/aic.16036}}.

\bibitem{Penn.CEJ2020}
A.~Penn, C.~Boyce, K.~Pruessmann, C.~Müller, Regimes of jetting and bubbling
  in a fluidized bed studied using real-time magnetic resonance imaging,
  Chemical Engineering Journal 383 (2020) 123185.
\newblock \href {https://doi.org/https://doi.org/10.1016/j .cej.2019.123185}
  {\path{doi:https://doi.org/10.1016/j .cej.2019.123185}}.

\bibitem{Penn.2019}
A.~Penn, C.~M. Boyce, N.~Conzelmann, G.~Bezinge, K.~P. Pruessmann, C.~R.
  M{\"u}ller, Real-time magnetic resonance imaging of fluidized beds with
  internals, Chemical Engineering Science 198 (2019) 117--123.
\newblock \href {https://doi.org/10.1016/j.ces.2018.12.041}
  {\path{doi:10.1016/j.ces.2018.12.041}}.

\bibitem{Yang.2003}
W.-C. Yang, Handbook of fluidization and fluid-particle systems, China
  Particuology 1~(3) (2003) 137.
\newblock \href {https://doi.org/10.1016/S1672-2515(07)60126-2}
  {\path{doi:10.1016/S1672-2515(07)60126-2}}.

\bibitem{Stehling.1991}
M.~K. Stehling, R.~Turner, P.~Mansfield, Echo-planar imaging: magnetic
  resonance imaging in a fraction of a second, Science (New York, N.Y.)
  254~(5028) (1991) 43--50.
\newblock \href {https://doi.org/10.1126/science.1925560}
  {\path{doi:10.1126/science.1925560}}.

\bibitem{Pruessmann1999}
K.~P. Pruessmann, M.~Weiger, M.~B. Scheidegger, P.~Boesiger, Sense: Sensitivity
  encoding for fast mri, Magnetic Resonance in Medicine 42~(5) (1999) 952--962.
\newblock \href
  {https://doi.org/https://doi.org/10.1002/(SICI)1522-2594(199911)42:5<952::AID-MRM16>3.0.CO;2-S}
  {\path{doi:https://doi.org/10.1002/(SICI)1522-2594(199911)42:5<952::AID-MRM16>3.0.CO;2-S}}.

\bibitem{Clift.2013}
R.~Clift, J.~R. Grace, M.~E. Weber, Bubbles, drops, and particles, {Dover
  Publ}, Mineola, NY, 2013.

\bibitem{Caicedo.2003}
G.~R. Caicedo, J.~J. Marqu{\'e}s, M.~G. Ru{\i}́z, J.~G. Soler, A study on the
  behaviour of bubbles of a 2d gas--solid fluidized bed using digital image
  analysis, Chemical Engineering and Processing: Process Intensification 42~(1)
  (2003) 9--14.
\newblock \href {https://doi.org/10.1016/S0255-2701(02)00039-9}
  {\path{doi:10.1016/S0255-2701(02)00039-9}}.

\bibitem{Boyce.2019}
C.~M. Boyce, A.~Penn, M.~Lehnert, K.~P. Pruessmann, C.~R. M{\"u}ller, Magnetic
  resonance imaging of single bubbles injected into incipiently fluidized beds,
  Chemical Engineering Science 200 (2019) 147--166.
\newblock \href {https://doi.org/10.1016/j.ces.2019.01.047}
  {\path{doi:10.1016/j.ces.2019.01.047}}.

\bibitem{Briens.1997}
L.~A. Briens, C.~L. Briens, A.~Margaritis, S.~L. Cooke, M.~A. Bergougnou,
  Characterization of channelling in multiphase systems. application to a
  liquid fluidized bed of angular biobone particles, Powder Technology 91~(1)
  (1997) 1--9.
\newblock \href {https://doi.org/10.1016/S0032-5910(96)03206-8}
  {\path{doi:10.1016/S0032-5910(96)03206-8}}.

\bibitem{Li.2014}
Y.~Li, H.~Fan, X.~Fan, Identify of flow patterns in bubbling fluidization,
  Chemical Engineering Science 117 (2014) 455--464.
\newblock \href {https://doi.org/10.1016/j.ces.2014.07.012}
  {\path{doi:10.1016/j.ces.2014.07.012}}.

\end{thebibliography}

\end{document}